\documentstyle[preprint,aps]{revtex} 
\begin{document} 
\draft 
 
 
\title{Triplet Waves in a Quantum Spin Liquid} 
\author{Guangyong Xu$^1$, C. Broholm$^{1,2}$, Daniel H. Reich$^1$, 
and M. A. Adams$^3$} 
\address{ 
$^1$Department of Physics and Astronomy, The Johns Hopkins 
University,\\ Baltimore, Maryland 21218\\ 
$^2$NIST Center for Neutron Research, National Institute of Standards 
and Technology, Gaithersburg, Maryland 
20899\\ 
$^3$ISIS Facility, CCLRC, RAL, Chilton, Didcot, Oxon, OX11 0QX, United Kingdom. 
} 
\date{\today} 
\maketitle 

\begin{abstract} 
We report a neutron scattering study of 
the spin-1/2 alternating bond antiferromagnet $\rm Cu(NO_3)_2\cdot
2.5D_2O$ for $0.06<k_BT/J_1<1.5$. For $k_BT/J_1 \ll 1$ the
excitation spectrum is dominated by a coherent singlet-triplet mode
centered at $J_1=0.442(2)$ meV with sinusoidal dispersion and a
bandwidth of $J_2=0.106(2)$ meV. A complete description of the
zero temperature contribution to the scattering function from this mode is
provided by the Single Mode Approximation. At finite temperatures we
observe exponentially activated band narrowing and damping.  The
relaxation rate is thermally activated and wave vector dependent with the periodicity of the reciprocal lattice.
 
\end{abstract} 
 
\pacs{75.10.Jm, 75.40.Gb, 75.50.Ee} 
 
\newpage
 
\narrowtext 
Transverse phonons and spin waves are propagating small amplitude
oscillations of a static order parameter in a broken symmetry phase.
Isotropic quantum antiferromagnets with a gap in their excitation
spectra can also support coherent wave-like excitations, but these
differ from phonons and spin waves in that they move through a system
with no static order. Specific examples of such systems include the
spin-1 chain\cite{affleck}, even-leg ladders\cite{Dagotto} and the
alternating bond spin-1/2 chain\cite{bonner0,bonner,vopo}.  Since they are
not based on the existence of a static order parameter that sets in at
a well defined transition temperature, coherent excitations in these
systems are expected to emerge smoothly with decreasing temperature as
short-range correlations develop.

In this letter we document this unique cooperative behavior through  
an experimental study of the 
temperature dependence of magnetic 
excitations in an isotropic, order-parameter-free quantum magnet. 
Specifically, we have studied magnetic excitations in the alternating 
spin-1/2 chain Copper Nitrate (CN) as a function of wave vector, energy, and   
temperature. The spin Hamiltonian for this system can  
be written\cite{bonner0,bonner} 
\begin{equation} 
\label{Hamiltonian} 
{\cal H}=\sum_n(J_1{\bf S}_{2n}\cdot{\bf S}_{2n+1}+  
J_{2} {\bf S}_{2n+1}\cdot{\bf S}_{2n+2}). 
\end{equation} 
Because $J_{2}/J_{1}\approx 0.24$ is small\cite{bonner,Eckert,Diederix}, 
it is useful to think of CN as a chain of pairs of 
spins-1/2.  Each pair has a singlet ground  
state separated from a triplet at $J_1\approx 0.44$  
meV. The weak inter-dimer coupling ($J_{2} \approx 0.11$ meV)  
yields a collective singlet ground state at low temperatures with triplet excitations 
that propagate coherently along the chain. 
We have characterized dynamic spin correlations 
in the temperature range $0.06<k_B T/J_{1}<1.5$, in considerable 
detail. We find that heating yields thermally activated band 
narrowing and  
an increased relaxation rate that varies with wave  
vector transfer with the periodicity of the reciprocal lattice. 
 
CN ($\rm Cu(NO_3)_2\cdot 2.5D_2O$), is monoclinic\cite{structure}  (space group $\rm I12/c1$) with low
$T$ lattice parameters $a=16.1$ \AA\ , $b=4.9$ \AA\ , $c=15.8$
\AA\ , and $\beta = 92.9^{\circ}$.  
The vector connecting dimers center to center 
is ${\bf u}_0=[111]/2$ for half the chains, and ${\bf
u}^{\prime}_0=[1\bar{1}1]/2$ for the other half.  The corresponding 
intra-dimer vectors are ${\bf d}_{1}=[0.252,\pm 0.027,0.228]$ respectively. In our experiment, the wave vector
transfer $\bf Q$ was perpendicular to $\bf b$ so the two sets of chains
contributed equally to magnetic neutron scattering.  The
sample  consisted of four 92\% deuterated,
co-aligned  single crystals with total mass 14.09 g\cite{GXuThesis}.

Inelastic neutron scattering measurements were performed on the inverse
geometry time of flight spectrometer IRIS at the Rutherford Appleton
Laboratory, UK \cite{IRISrefs}. Disk choppers selected an incident
spectrum from 1.65 meV to 3.25 meV pulsed at 50 Hz, and a backscattering
pyrolytic graphite analyzer bank selected a final energy $E_f=1.847$
meV. The Half Width at Half Maximum (HWHM) elastic energy
resolution was $10.5~\mu$eV.  The $\bf b $
direction of CN was perpendicular to the horizontal
scattering plane and [101] pointed towards the low angle part
of the  analyzer bank at an angle of $\phi=20(1)^\circ$ to the direct beam.
The direction for rotation of the $\bf a$ axis into
the $\bf c$ axis coincided with the direction of decreasing scattering
angle.  In this configuration the projection of wave vector transfer
on the chain $Q_{\parallel}=k_i\cos\phi-k_f\cos (\phi-2\theta)$ takes
on a unique value for each detector in the range of scattering angles
$20^{\circ}<2\theta <160^{\circ}$ covered. 
We can therefore present our data  as a function of energy transfer $\hbar\omega$ and wave vector transfer
along the chain  $\tilde{q}={\bf Q}\cdot{\bf u}_0$. 
In addition there is a 
specific value of $Q_{\perp}=k_i\sin\phi-k_f\sin (\phi-2\theta )$ associated
with each $(\tilde{q},\hbar\omega )$ point such that sensitivity
to dispersion perpendicular to the chain is maintained in the projection.  
Count-rates were normalized to incoherent elastic
scattering from the sample to provide absolute
measurements of $ \tilde{I}({\bf Q},\omega)=|\frac{g}{2}F(Q)|^2 2 {\cal
S}({\bf Q},\omega )$.  Here $g= \sqrt{(g_{b}^2 + g_{\perp}^2)/2}=2.22$
\cite{bonner}, $F(Q)$ is the magnetic form factor for Cu$^{2+}$
\cite{formf}, and ${\cal S}({\bf Q},\omega )$ is the scattering function\cite{Lovesey}.

Figures \ref{Figure1}(a)-(c) show the measured neutron scattering
spectrum at $T=0.3$ K, 2 K, and 4 K. Focusing at first 
on the 0.3 K data,
we observe a resonant mode centered around $J_1=0.44$
meV with bandwidth $\approx J_{2}$, consistent with the predictions of
perturbation theory \cite{HarrisTennant}. The mode energy has the periodicity
$2\pi$ of the one-dimensional reciprocal lattice with minima
for $\tilde{q}={\bf Q\cdot u_0}=n\cdot 2\pi$, indicating antiferromagnetic inter-dimer interactions. The intensity of the mode varies with a periodicity
that is incommensurate with that of the dispersion relation. 
There is a simple explanation for this, namely that the intra-dimer spacing that enters in the neutron scattering cross section is incommensurate with the period of the alternating spin chain\cite{Eccleston98}. 
An exact sum-rule for ${\cal S}({\bf Q},\omega )$ provides the following direct link between the microscopic structure 
and the intensity of inelastic neutron scattering\cite{hohenberg}:
\begin{eqnarray} 
\label{fmom} 
\hbar \langle \omega\rangle_{\bf Q}&\equiv&
\hbar^2\int_{-\infty}^{\infty}\omega  {\cal S}({\bf Q},\omega )d\omega \nonumber \\
&=& -\frac{2}{3} 
\sum_{\bf d}J_{\bf d}\langle{\bf S}_0\cdot{\bf S}_{\bf d}\rangle 
(1-\cos {\bf Q\cdot d}), 
\end{eqnarray}
where \{{\bf d}\} is the set of all bond vectors connecting a spin to its neighbors.
Figure 2 (a) shows $\hbar \langle \omega\rangle_{\bf Q}$ at $T=0.3$ K and $T=4$ K derived by integrating the corresponding data sets in Fig. 1. 
The solid lines show fits based on Eq.~\ref{fmom} including only
intra-dimer 
correlations $\langle {\bf S_0 \cdot S_{d_1}} \rangle$ and a constant to account
for multiple scattering (see below).  
The  excellent agreement between model and data provides direct evidence for singlet
formation between spins separated by $\bf d_1$. The modulation amplitude is $J_1\langle {\bf S_0 \cdot S_{d_1}} \rangle$.

From the wave vector dependence of energy integrated intensities we turn to spectra at fixed wave vector transfer. Figure 3 shows cuts through raw data for $\tilde{q}=2\pi$ and $\tilde{q}=3\pi$ at three temperatures. At $T=0.3$ K we see resolution limited peaks. Upon increasing temperature these peaks broaden and shift towards the center of the band. Careful inspection also reveals that the higher energy peak at $\tilde{q}=3\pi$ broadens less than the lower energy peak at $\tilde{q}=2\pi$. From gaussian fits to constant energy cuts such as these, we extract the resonance energy and half width at half maximum versus wave vector transfer shown in Fig. 2 (c) and (d). This analysis shows that the finite $T$ relaxation rate is wave vector dependent with the apparent periodicity of the reciprocal lattice.

We now proceed to extract the detailed temperature dependence of the integrated intensity, the effective bandwith and the relaxation rate. To take full advantage of the wide sampling of ${\bf Q}-\omega$ space in our experiment and to account for resolution effects, the analysis is based on ``global fits'' of the following phenomenological form for ${\cal S}({\bf Q},\omega )$ 
to the complete ${\bf Q}$ and $\omega$ dependent data set at each temperature. For $\hbar\omega >0$ we write
\begin{equation} 
\label{sqw} 
{\cal S}({\bf Q},\omega)=\frac{\hbar \langle \omega \rangle_{\bf Q}}{\epsilon ({\bf Q})}\frac{1}{1-\exp (-\beta \epsilon ({\bf Q}))}
f(\hbar\omega-\epsilon({\bf Q})),
\end{equation}
where $f(E)$ is a normalized spectral function. The other terms implement the first moment sum-rule of Eq.~\ref{fmom} in the limit where $f(E)$ is sharply peaked
on the scale of $J_1$. Eq.~\ref{sqw} represents the ``Single Mode Approximation'' (SMA) that has been used with success to link the equal time structure factor and the dispersion relation for collective modes in numerous many body systems\cite{hohenberg,girvin,ma92}. Given that two magnon scattering carries less than 1\% of the spectral weight for CN\cite{TennantUnpub}, the SMA should be excellent at sufficiently low $T$. For the dispersion relation we use the following variational form 
based on first order perturbation theory\cite{HarrisTennant}  
\begin{equation} 
\label{dispersion1} 
\epsilon ({\bf Q})=J_1-\frac{1}{2}\sum_{\bf u}J_{\bf u}\cos {\bf  
Q\cdot u} , 
\end{equation} 
The vectors \{{\bf u}\} connect neighboring  
dimers center to center, both 
within and between the chains. For comparing to the experimental data, ${\cal S}({\bf Q},\omega)$ was convolved with the instrumental resolution\cite{IRISrefs}. In all fits it was necessary to take into  
account multiple neutron scattering events involving elastic 
incoherent  
nuclear scattering  followed or preceded by coherent inelastic  
magnetic scattering. We believe that 
such processes are responsible for the weak 
horizontal band of intensity that is visible
in Fig.~\ref{Figure1}(a).  For $T=0.3$ K we used $f(E)=\delta (E)$ and  
obtained the fit shown in Fig 1 (d). To better evaluate the quality of the global fit we also show cuts through the model calculation as solid lines in Fig. 3 (a). There is excellent agreement between model and data with an overall 
prefactor and four exchange constants as the only fit parameters.  The prefactor refined to $\langle {\bf S_0 \cdot S_{d_1}} \rangle=-0.9(2)$ consistent with the value of -3/4 expected for isolated singlets. Allowing for inter-dimer correlations by including the corresponding term from 
Eq.~\ref{fmom} in the global fit yields 
$\langle {\bf S}_0 \cdot {\bf S}_{{\bf d}_2}\rangle=-0.04(8)$.  
From the best fit parameters in the variational dispersion relation 
we get $J_1=0.442(2)$ meV and 
$J_2=0.106(2)$ meV  
for the two intra-chain interactions and 
$J_L=0.012(2)$ meV and $J_R=0.018(2)$ meV for dimers separated by 
${\bf u}_{L} = [\frac{1}{2},0,0]$ and ${\bf u}_{R} =  
[0,0,\frac{1}{2}] $ respectively. Because $\bf b\perp Q$ throughout the 
experiment our data do not yield an estimate for inter-chain interactions between dimers displaced by $[1\mp 1 1]/2$.  However, in a separate measurement with the crystal oriented in the $(hkh)$ plane\cite{stone}, we were able to place an upper limit of 0.02 meV on the corresponding parameter in Eq.~\ref{dispersion1}.
The value $J_{2}/J_{1} = 0.240(5)$ that we obtain is  measurably
smaller than the value $J_{2}/J_{1} \approx 0.27$\cite{bonner} derived  
from magnetic susceptibility data. A likely explanation for this 
discrepancy is  
that susceptibility measurements cannot distinguish 
intra-chain from inter-chain interactions. 

To analyze the finite $T$ data we replaced the spectral function with a normalized Gaussian with HWHM
\begin{equation} 
\Gamma (\tilde{q})=\Gamma_0+\frac{\Gamma_1}{2}\cos\tilde{q}, 
\end{equation}
The functional form for the dispersion relation (Eq.  
\ref{dispersion1}) was maintained, but to account for the bandwidth 
narrowing that  
is apparent in the raw data, we introduced an overall 
renormalization  
parameter relating finite $T$ ``effective'' exchange constants in 
the  
variational dispersion relation to the bare temperature independent  
exchange parameters: $\tilde{J}_{\bf u}=n(T)J_{\bf u}$.  
The fits obtained are excellent as can be ascertained by  
comparing the left and right columns  
in Fig.~\ref{Figure1} and the solid lines through the data in  
Fig.~\ref{Figure3}.  The solid lines in Fig.~\ref{Figure2}(b) and (c) show the
dispersion relation and $\tilde{q}$-dependent relaxation rate derived from the global fits. They are consistent with the data points derived from constant-$\tilde{q}$ cuts indicating that the variational forms employed for $\epsilon (\tilde{q})$ and $\Gamma (\tilde{q})$ have not biased the global fitting analysis. We note that the quality of 
the fits did not change significantly when using a Lorentzian 
rather than a Gaussian spectral function. So while the data provide reliable information on the $\tilde{q}$- and $T$-dependent relaxation rate, they do not accurately determine the spectral function.

The temperature-dependent parameters derived from this analysis are 
shown in Fig.~\ref{Figure4}.  
The prefactor for global fits at each temperature yields the intra-dimer spin correlation function, which we plot versus $T$ in Fig.~\ref{Figure4}(a). As expected $|\langle {\bf S}_0 \cdot {\bf S}_{{\bf d}_1}\rangle|$  decreases with increasing 
$T$ as the populations of the four states of each spin pair equalize. For an isolated 
spin pair it can be shown that  
$\langle {\bf S}_0 \cdot {\bf S}_{\bf 
d}\rangle=-(3/4)\Delta n(\beta J_1)$, where $\Delta n(\beta J_1)=(1-e^ {-\beta J_1})/(1+3e^ {-\beta J_1})$ is the singlet triplet population difference. After fitting a scale factor, this form provides an excellent
description of the temperature dependence of the  
data (solid line in  
Fig.~\ref{Figure4}(a)). The Random Phase Approximation (RPA) applied to interacting spin dimers\cite{rpa1,rpa2,rpa3,rpa4} predicts that 
the bandwidth renormalization factor, $n(T)$, also follows
the singlet triplet population 
difference. A similar result holds when the singlet ground state is induced by single ion anisotropy\cite{lindgaard}.
We compare $n(T)$ to $\Delta n(\beta J_1)$ in 
Fig.~\ref{Figure4}(b). While there is qualitative agreement, the RPA 
clearly predicts more bandwidth narrowing than is actually observed. 
 
Triplet relaxation is due to scattering from
the thermal ensemble of excited states. Because the density of triplets is thermally activated, the relaxation rates should be too.  
Fits of a simple activated form $\Gamma_{i} =  
\gamma_i \exp(-\Delta_i/k_{B}T)$ to the $T\le 4$ K data in Figs.~\ref{Figure4}(c) and 
(d)  give $\Delta_0=0.24(2), \gamma_0=0.10(2)$ meV, $\Delta_1=0.32(3)$, and 
$\gamma_1=0.08(2)$ meV.   
If we instead fix $\Delta_i \equiv J_1$ and allow for a power law prefactor
: $\Gamma_i (T)= \gamma_i(J_1 /k_BT)^{\alpha_i}\exp
(-J_1/k_BT)$ we also obtain excellent fits
with $\alpha_0 =1.0(2), \gamma_0=0.13(4)$ meV,
$\alpha_1=0.6(2)$,
and $\gamma_1=0.10(1)$ meV.

In summary, we have examined dynamic spin correlations in the strongly
alternating spin chain $\rm Cu(NO_3)_2\cdot 2.5D_2O$ for
$0.06<k_BT/J_1<1.5$. For $k_BT \ll J_1$ we find a coherent dispersive
triplet mode whose contribution to the scattering
function is perfectly accounted for by the SMA, and we have determined
accurate values for inter-dimer interactions in the
material.  Upon heating, the sharp dispersive mode gradually
deteriorates through band-narrowing and the development of a
wave-vector dependent lifetime. A semiclassical theory for finite-$T$
excitations in gapped spin chains was recently developed by Sachdev and
Damle\cite{sachdev}. It relies on $\Delta /J$ being a small parameter
as is the case for the Haldane phase of spin-1 chains, and in weakly
dimerized spin-1/2 chains. In CN the spin gap is instead
much greater than the magnetic bandwidth. Our data should provide a
focus for theoretical attempts to describe finite temperature
properties in this ``strong coupling" limit.
 
It is a pleasure to acknowledge the assistance provided by the staff of
the RAL during the measurements, and we thank R. Eccleston for
illuminating discussions. We also thank W. Wong-Ng for help on
characterizing crystals, and R. Paul for neutron activation analysis at
NIST.  Work at JHU was supported by the NSF through DMR-9453362 and
DMR-9357518.  DHR acknowledges support from the David and Lucile
Packard Foundation.

\begin{figure} 
\caption{Normalized scattering intensity, 
$\tilde{I}(\tilde{q},\omega)$, for 
$\rm Cu(NO_3)_2\cdot 2.5D_2O$ at $T = 0.3 $ K (a), $T = 2$ K (b), 
and $T = 4$ K (c). Maximum intensity on the color scale shown above
are $\tilde{I}_{max}=28$ meV$^{-1}$, 16 meV$^{-1}$ and 8 meV$^{-1}$ respectively. (d)-(f) 
show model calculations based on Eq.~\ref{sqw}} 
\label{Figure1} 
\end{figure}

\begin{figure} 
\caption{First moment computed from Fig. 1 (a) and (c) using $\hbar\langle\omega\rangle_{\bf Q}=\hbar^2\int_0^{\infty}\omega (1-e^{-\beta\hbar\omega}) {\cal S}({\bf Q},\omega )d\omega$
 (a), 
resonance energy (b), and HWHM $\delta E(\tilde{q})$ (c) of the triplet mode in copper 
nitrate  
 at $T=0.3$ K and $T=4$ K. The points in (b) and (c) were obtained 
from fits to constant- $\tilde{q}$ cuts through the data in Fig. 1. 
Dashed line in (c) shows the HWHM instrumental energy  
resolution at $\hbar\omega=0.45$ meV. 
Solid lines in (b) and (c) were obtained from global fits at each $T$. } 
\label{Figure2} 
\end{figure}

\begin{figure} 
\caption{Constant-$\tilde{q}$ cuts through the  
data shown in Fig. 1 at $\tilde{q}=2\pi$ and $3\pi$. The solid  
lines are the results of  global fits at each $T$. Dashed lines  
indicate 
peak positions at $T = 0.3$ K, and highlight band narrowing at 
higher $T$.} 
\label{Figure3} 
\end{figure}

\begin{figure} 
\caption{Parameters characterizing the  
temperature dependence of dynamic spin correlations in  
CN, 
obtained from fits shown in Figs. 1 and 3. 
The solid lines in (a) and (b) show RPA theory. The solid lines in 
(c) and (d) are fits to $\Gamma_i (T)= \gamma_i(J_1 /k_BT)^{\alpha_i}\exp
(-J_1/k_BT)$. } 
\label{Figure4} 
\end{figure}
 
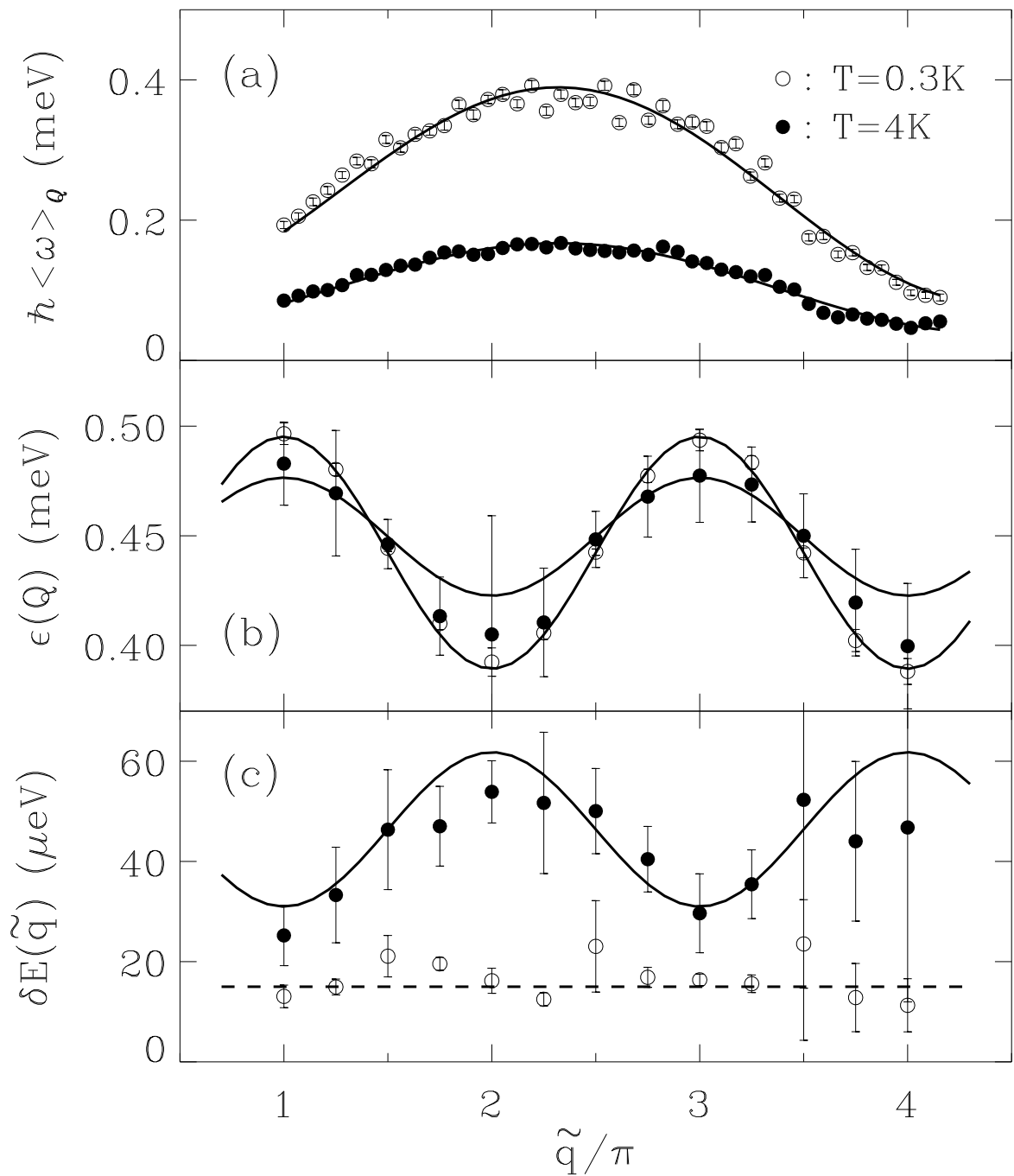
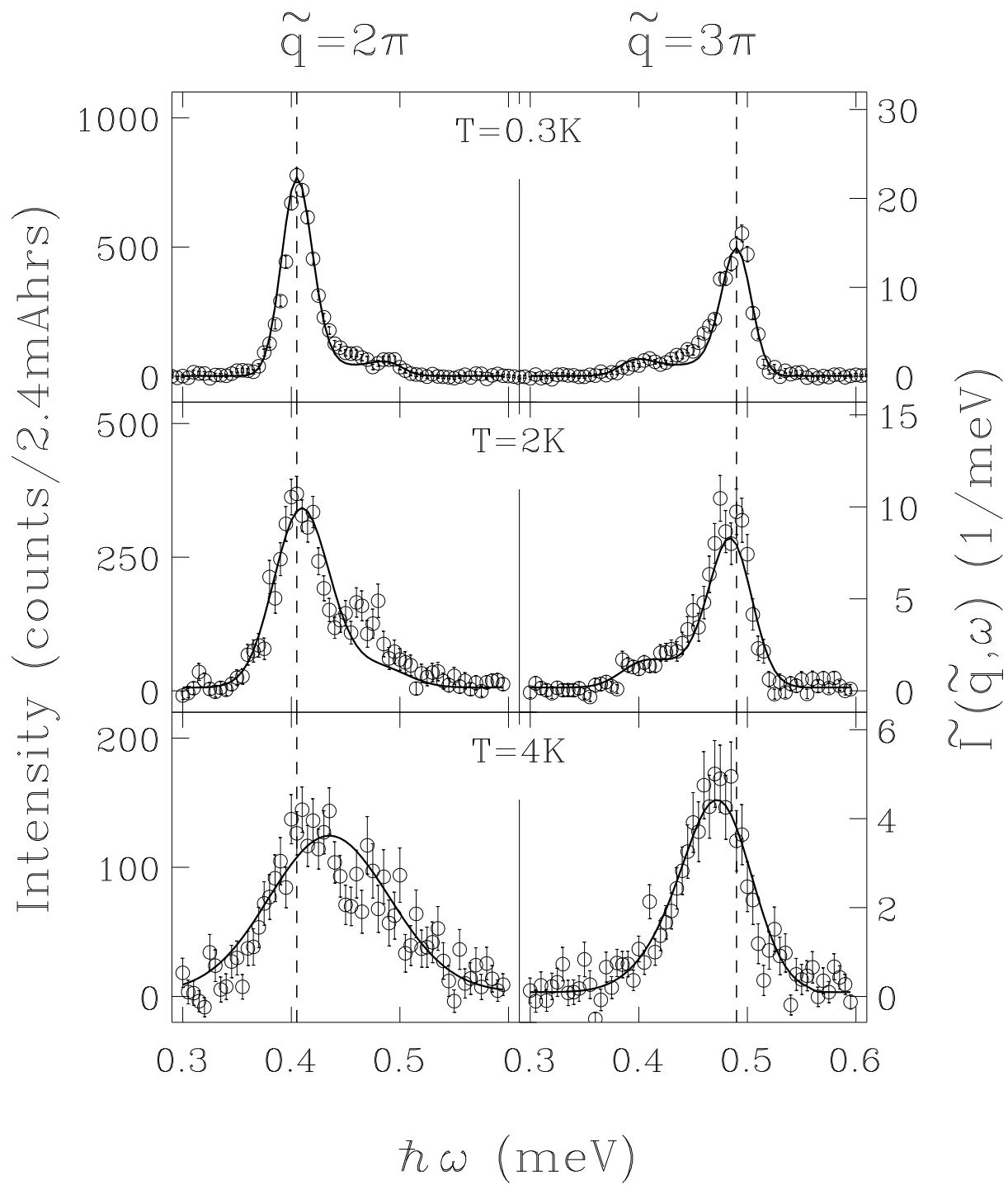
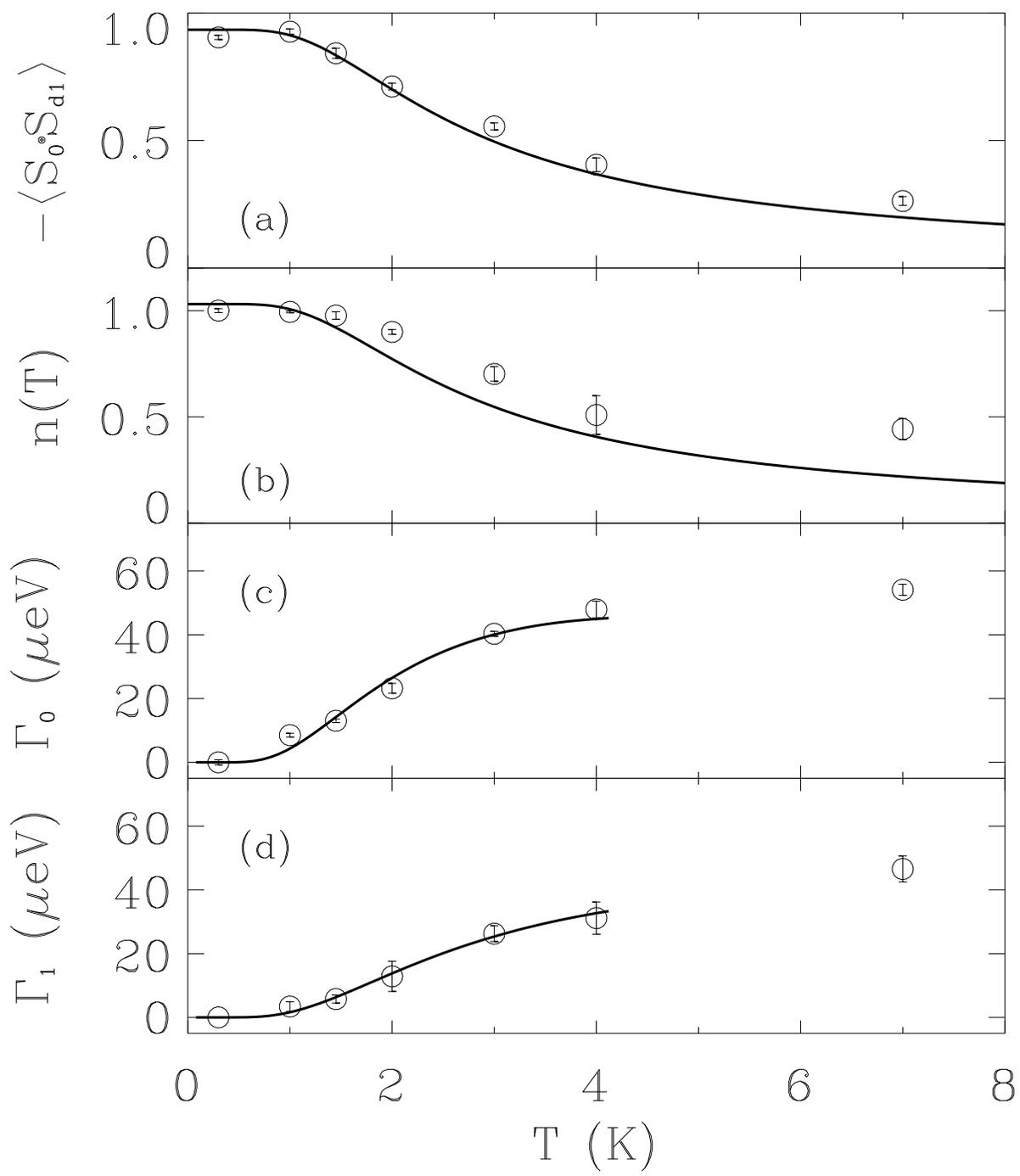
\end{document}